\documentclass[11pt,twoside]{article}

\usepackage{asp2006}

\begin{document}
\title{On magnetic fields in broad-line blazars}
\author{Rafa{\l} Moderski and Marek Sikora}
\affil{Nicolaus Copernicus Astronomical Center, Bartycka 18, 00-716
  Warsaw, Poland}

\begin{abstract} 
High energy spectra of broad-line blazars can be reproduced by both
synchrotron-self-Compton (SSC) models and external-Compton (EC)
models.  However, as is known from numerical modeling, SSC scenarios
require much weaker magnetic field than EC ones.  In this paper we
quantify these results analytically.  We show that for blazars with $q
\equiv F_{\gamma}/F_{syn} \gg 1$ the SSC models predict a
magnetic-to-electron energy density ratio lower than $1/q^2$, and that
EC models allows equipartition between magnetic fields and electrons
for any high $q$ provided $\Gamma \sim 16 (q/10)^{1/2}
(F_{syn}/F_{d})^{1/4}$, where $\Gamma$ is the Lorentz factor of the
source and $F_{d}$ is the radiation flux from the accretion disk.
\end{abstract}

\section{Introduction}
Blazars -- the AGNs dominated by the Doppler boosted radiation of the
relativistic jets -- provide an exceptional opportunity for studies of
physics of innermost parts of AGN jets. However, to take advantage of
this opportunity the radiative mechanisms of the jet emission must be
identified.  While a low-energy component of blazar spectra is
uniquely identified as due to the synchrotron process, the nature of a
high-energy component is still uncertain.  It can be contributed by
inverse-Compton (IC) radiation of directly accelerated electrons
(hereafter by electrons we mean both electrons and positrons), by
synchrotron radiation of pair cascades powered by hadronic processes
and accompanying pair cascades, and, finally, by synchrotron emission
of protons and muons \citep[see review by][]{sm01}.  Seed photons for
the IC process are provided by locally operating synchrotron mechanism
as well as by external sources, like broad line region (BLR) and/or
dusty tori.  The IC models involving Comptonization of local
synchrotron radiation are called synchrotron-self-Compton (SSC)
models, and those involving Comptonization o external radiation are
usually coined external-Compton (EC) models.

In this paper we focus on blazars with a dense radiative environment,
i.e. those hosted by quasars.  Their $\gamma$-ray fluxes usually
exceed fluxes of a low-energy, synchrotron spectral component by a
large factor, and their X-ray spectra are often very hard.  About 30\%
of them have X-ray spectral index $\alpha_X < 0.5$
\citep{cap97,rt00,dsg05}.  Such spectra cannot be produced in a fast
cooling regime.  This practically eliminates models which predict
production of X- and $\gamma$-rays by synchrotron emission of
ultrarelativistic electrons, and also put severe constraints on SSC
models.  We demonstrate that spectra of such blazars can be explained
in terms of the SSC models only for magnetic fields much below the
equipartition with electrons and that no such constraint applies for
EC models provided a bulk Lorentz factor of the source is sufficiently
high.

\section{Magnetic fields in SSC and ERC models}
We investigate here broad-line blazars (those hosted by quasars) with
$q \equiv F_{\gamma}/F_{syn}\gg 1$ and assume that their $\gamma$-ray
component is produced by IC process.  Both components are assumed to
be produced by the same population of electrons which in the source
co-moving frame have the isotropic distribution. Since in the
broad-line blazars the KN and pair production effects are likely to be
insignificant \citep{mod05}, we ignore them in our considerations
below.  With all these assumptions
\begin{equation}
u_{\gamma}'/u_{seed}' = u_{SSC}'/u_{syn}' = u_{syn}'/u_B' =
u_{EC}'/u_{ext}' = A \,
\label{eq:ugus}
\end{equation}
where $A$ is the Thomson amplification factor, $u_B'$ is the magnetic
energy density, and $u_{\gamma}'$, $u_{seed}'$, $u_{syn}'$,
$u_{SSC}'$, $u_{ext}'$, $u_{EC}'$, are energy densities of
$\gamma$-ray, seed photon, synchrotron, SSC, external and EC radiation
fields, respectively, all as measured in the source co-moving frame.

\subsection{Magnetic vs. radiation energy density}
In the SSC model $u_{seed}'=u_{syn}'$ and $u_{\gamma}' = u_{SSC}'$.
Therefore,
\begin{equation}
q = {F_{\gamma} \over F_{syn}} = {u_{SSC}'\over u_{syn}'} = A \, ,
\label{eq:qssc}
\end{equation}
and
\begin{equation}
{u_B' \over u_{\gamma}'} = {u_{syn}' \over u_{SSC}'} {u_B' \over
u_{syn}'} = {1 \over A^2} = {1 \over q^2} \, .
\label{eq:ussc}
\end{equation}

In the EC model $u_{seed}'=u_{ext}'$ and $u_{\gamma}' = u_{EC}'$, and,
therefore,
\begin{equation}
q = {F_{\gamma} \over F_{syn}} = {f u_{EC}'\over u_{syn}'} \, ,
\label{eq:qerc}
\end{equation}
and 
\begin{equation}
{u_B' \over u_{\gamma}'} = {u_{syn}'/A \over u_{EC}'} = {f \over
Aq} \, ,
\label{eq:uerc}
\end{equation}
where the factor $f \simeq ({\cal D}/\Gamma)^2$ accounts for
anisotropy of IC radiation in the source comoving frame \citep{msb03}
and ${\cal D} \equiv 1/[\Gamma(1-\beta \cos \theta_{obs})]$ is the
Doppler factor. Note that here, in general, $q \ne A$.

Hence, in SSC models $q \gg 1$ is possible only for $u_B' \ll
u_{\gamma}'$, while EC can give $q \gg 1$ even for $u_B' >
u_{\gamma}'$, provided $A < f/q$.

\subsection{Electron vs. radiation energy density}
Let us approximate the geometry of the source as a piece of a
cylinder, of a cross-sectional radius $R$ and height $\lambda'$ as
measured in the source co-moving frame.  An amount of energy
acumulated in relativistic electrons during the time period of the
injection of relativistic electrons is
$$ E_e' = (1-\eta_{rad}) L_{e,inj}' t_{inj}' \, , \eqno(6) $$
and energy density of the relativistic electrons is
$$ u_e' = {E_e' \over \pi R^2 \lambda'} = {(1-\eta_{rad}) L_{e,inj}'
t_{inj}' \over \pi R^2 \lambda'} \, , \eqno(7) $$
where $L_{e,inj}$ is the the rate of the energy injection via
acceleration of relativistic electrons, $t_{inj}'$ is the injection
time scale, and $\eta_{rad}$ is the fraction of the total electron
energy converted via radiative processes to photons.  Energy of the
radiation produced during the flare is $E_{rad}' =
\eta_{rad}L_{e,inj}' t_{fl}' $, and energy density of the emitted
radiation is
$$ u_{rad}' = {\eta_{rad} L_{e,inj}' \over 2 \pi R^2 c} \, , \eqno(8)
$$
provided $\lambda' < R$. Hence,
$$ {u_{rad}' \over u_e'} = {\kappa \, \eta_{rad} \over 1-\eta_{rad}}
 \, .  \eqno(9) $$
where $\kappa = \lambda'/ (2 t_{inj}' c)$.

\subsection{Magnetic vs. electron energy density}
Noting that for $q \gg 1$, $u_{rad}' \simeq u_{\gamma}'$, and
combining equations (3), (5), and (9), we obtain for SSC model
$$ {u_B' \over u_e'} = {1 \over q^2} \, {\kappa \, \eta_{rad} \over
1-\eta_{rad}} \, ,\eqno(10) $$
and for EC model
$$ {u_B' \over u_e'} = {f \over Aq} \, {\kappa \, \eta_{rad} \over
1-\eta_{rad}} \, . \eqno(11) $$
In the broad-line blazars radiation efficiency $\eta_{rad}$ is
typically of the order of $0.5$, and then, for $\lambda' < 2 c
t_{inj}'$, the SSC models of the high-q blazars require magnetic
fields much below equipartition, $u_B'/u_e' < 1/q^2$.

In the case of EC models, equipartition of magnetic fields with
electrons is possible for any $q$, but requires
$$ A \sim {f \over q} \, {\kappa \, \eta_{rad} \over 1-\eta_{rad}} \,
. \eqno(12) $$
Noting that
$$ u_{EC}' = {L_{\gamma}' \over 2 \pi R^2 c} = {2 F_{\gamma} \over c}
\left(d_L \over R \right)^2 {1 \over f {\cal D}^4} \, \eqno(13) $$
and
$$ u_{ext}' = \Gamma^2 \, {\xi L_d \over 4 \pi r^2 c} = \Gamma^2 {\xi
F_d \over c} \left(d_L \over r\right)^2 \, , \eqno(14) $$
we obtain 
$$ A={u_{EC}' \over u_{ext}'} = {2 F_{\gamma} \over \xi F_d} \, {1
\over f {\cal D}^4 (\Gamma \theta_j)^2} \, , \eqno(15) $$
where $\xi$ is the fraction of the disk radiation reprocessed into
broad lines and/or IR dust radiation at a distance $r$ at which the
flare is produced, $\theta_j = R/r$, and $d_L$ is the luminosity
distance of the blazar. Combining equations (12) and (15) gives the
condition for the Doppler factor ${\cal D}$ to have $u_B' = u_e'$,
$$ {\cal D} = \left( 2 q (1 -\eta_{rad}) F_{\gamma} \over f^2
\eta_{rad} \kappa (\Gamma \theta_j)^2 \xi F_d \right)^{1/4} \,
. \eqno(16) $$
For ${\cal D} \sim \Gamma \sim 1/\theta_j$, $\eta_{rad} \sim 0.5$, and
noting that $F_{\gamma}/F_d = (F_{\gamma}/F_{syn})(F_{syn}/F_d)=q
(F_{syn}/F_d)$, we obtain
$$ \Gamma \sim 16 \, \left(q \over 10\right)^{1/2} \, \left( {\kappa
\over (\xi/0.3)}{F_{syn} \over F_d} \right)^{1/4} \, . \eqno(17) $$

\section{Conclusion}
Large $\gamma$-ray excesses ($q \gg 1$), observed in many broad-line
blazars can be produced by SSC process only if $B \ll B_{equip}$. No
such constraint applies to the ERC model, provided a jet Lorentz
factor is respectively high: $B \sim B_{equip}$ fits the model if
$\Gamma \sim (q/10)^{1/2} (F_{syn}/F_d)^{1/4}$.

\acknowledgements This work was partially supported by Polish MNiSW
grant 1P03D00928.



\begin{thebibliography}{}
\bibitem[Cappi et al.(1997)]{cap97}
Cappi, M., et al.~1997, \apj, 478, 492
\bibitem[Donato, Sambruna, \& Gliozzi(2005)]{dsg05}
Donato, D., Sambruna, R.M., \& Gliozzi, M.~2005, \aap, 433, 1163
\bibitem[Moderski, Sikora, \& B{\l}a{\.z}ejowski(2003)]{msb03}
Moderski, R., Sikora, M., \& B{\l}a\.zejowski, M.~2003, \aap, 406, 855
\bibitem[Moderski, et al.(2005)]{mod05}
Moderski, R., et al.~2005, \mnras, 364, 1488
\bibitem[Reeves \& Turner(2000)]{rt00}
Reeves, J.N., \& Turner, M.J.L.~2000, \mnras, 316, 234
\bibitem[Sikora, et al.(2005)]{sik05}
Sikora, M., et al.~2005, \apj, 625, 72
\bibitem[Sikora \& Madejski(2001)]{sm01}
Sikora, M., \& Madejski, G.M.~2001, in ``High Energy Gamma-Ray
Astronomy'', AIP Conference Proceedings 558 (eds F.A. Aharonian \&
H.J.~V\"olk), p.~275
\end{thebibliography}
\end{document}